\documentclass[12pt,a4paper]{article} 
\usepackage[dvipsnames,usenames]{color}
\usepackage{graphicx}
\usepackage[normalem]{ulem}
\usepackage{lscape}
\usepackage{multirow}

\begin{document}

\title{Dyson Spheres around White Dwarfs}

\author{\.{I}brahim 
Semiz\thanks{mail: ibrahim.semiz@boun.edu.tr} \hspace{2 mm} and
Salim O\u{g}ur\thanks{mail: salim.ogur@boun.edu.tr}\\
Bo\u{g}azi\c{c}i University, Department of Physics\\
Bebek, \.{I}stanbul, TURKEY}
    
\date{ }

\maketitle

\begin{abstract}
A Dyson Sphere is a hypothetical structure that an advanced civilization might build around a star to intercept all of the star's light for its energy needs. One usually thinks of it as a spherical shell about one astronomical unit (AU) in radius, and surrounding a more or less Sun-like star; and might be detectable as an infrared point source.

We point out that Dyson Spheres could also be built around white dwarfs. This type would avoid the need for artificial gravity technology, in contrast to the AU-scale Dyson Spheres. In fact, we show that parameters can be found to build Dyson Spheres suitable --temperature- and gravity-wise-- for human habitation. This type would be much harder to detect.
\end{abstract}


\section{Introduction}
\label{intro}

\begin{figure}[h]
\centering
\includegraphics[scale=0.8]{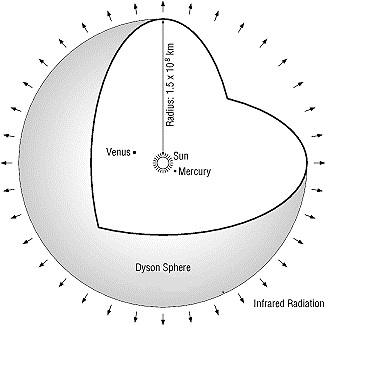}
\caption{A Dyson Sphere with 1 AU radius in Sol system. 
\footnotesize {\it Adapted from David Darling,} http://www.daviddarling.info/encyclopedia/D/Dysonsp.html, {\it acsessed 8 Dec. 2014.}}
\label{fig:DS1AU}
\end{figure}

The "Dyson Sphere" \cite{dyson_sphere} concept is well-known in discussions of possible intelligent life in the universe, and has even infiltrated popular culture to some extent, including being prominently featured in a {\em Star Trek} episode \cite{ST-TNG_relics}.  In its simplest version, it is a spherical shell that totally surrounds a star to intercept all of the star's light. If a Dyson Sphere (from here on, sometimes ``Sphere'', sometimes DS) was built around the Sun, e.g. with same radius (1 AU) as Earth's orbit (Fig. \ref{fig:DS1AU}),
it would receive all the power of the Sun, $3.8 \times 10^{26}$ W, in contrast to the power intercepted by Earth, $1.7 \times 10^{17}$ W. These numbers can be compared to the current power consumption of Earth, $1.7 \times 10^{13}$ W \cite{bp}, but one has to keep in mind that  this last figure does not include the power to keep Earth at $\sim$ 280 K rather than the $\sim$ 3 K of interstellar space, nor the energy we get from food to keep our body temperatures constant and our physiological processes running. Such a Dyson Sphere\footnote{A radius of 1 AU would seem to be chosen so that if we lived on that sphere, the Sun would look the same as it did from the surface of Earth. However, such a sphere would have a temperature of $\sim$400 K, since the infrared-emitting surface of the Earth is four times the  the sunlight-receiving cross-section, while for the Dyson Sphere, the two surfaces are equal.} also would represent a living area orders of magnitude larger than Earths's surface ($2.8 \times 10^{23}$ ${\rm m^{2}}$ vs. $5.1 \times 10^{14}$ ${\rm m^{2}}$).

The amount of non-physiological power at the disposal of the average individual has usually been considered as one of the measures of technological advancement in our recent history. Possibly extrapolating this trend, Kardashev \cite{Kardashev} has given a classification of hypothetical intelligent civilizations according to their energy use. A Type I civilization is able to harness all the energy resources of a planet, a Type II civilization can use all the energy resources of a star (and its planetary system), and a Type III civilization, all the energy resources of a galaxy (Obviously, our civilization is striving to become Kardashev Type I). With the transition from the industrial age to the information age, some \cite{Kardashev_crit_1,Kardashev_crit_2}  suggested that the measure of advancement should be information-processing capability, and other measures were also suggested, e.g. the ability to manipulate smaller scales (the ``Barrow scale'') \cite{Barrow}. Still, the Kardashev classification is part of the lore of discussions of intelligent life in the universe.

 A Dyson Sphere represents a particular realization of the Kardashev Type II stage, the obvious way to "catch" all the power of a star. Dyson \cite{dyson_sphere} in the original paper\footnote{He seems to have taken the idea from the science-fiction writer Olaf Stapledon \cite{disturbing,starmaker}, so maybe it should be called the Stapledon-Dyson Sphere.} points out that such a sphere would have to emit thermal radiation with total power\footnote{"luminosity" in astronomy parlance.} equal to that of the star. Since it would look like a point source as seen from astronomical distances, and other astronomical cold point-like objects would have much less luminosities than stellar values, it would stand out in astronomical IR images. 
 
The simplest form of the Dyson Sphere, a solid spherical shell, is problematic: It would be subject to unaccetably large stresses and its equilibrium around the star is neutral at best  (as opposed to stable, see Appendix). Therefore, variants were suggested where the ``sphere'' actually consists of pieces in independent orbits (A ``Dyson Swarm''). Another consideration is gravity: If the sphere were built in the Sol system with 1 AU radius, the gravity due to the Sun would be only $5 \times 10^{-4} \; g$, so humans could not live on it without either genetic modification to become compatible with microgravity, or a technology of artificial gravity. Genetic modification may be undesirable socially or ethically, and artificial gravity is obviously beyond the scope of current paradigms of physics, maybe forever impossible. One could rotate the Dyson Sphere to simulate gravity, but then only equatorial regions would be really usable, unless the sphere consists of rings at different latitudes, which rotate about the same axis, but with different angular velocities. Or one could just take the equatorial regions, i.e. a ring instead of a sphere, leading to the intermediate case  described in the novel {\em Ringworld} \cite{ringworld} and its sequels by Larry Niven. But the {\em Ringworld} is unstable (see Appendix).

In this work, we will mostly ignore the mechanical complications associated with the simplest case, the rigid spherical shell, but make the case for a variant where the central object is different: A white dwarf (WD). We show that a smaller Dyson Sphere built around a typical white dwarf can simultaneously satisfy the temperature and gravity requirements for human, and therefore presumably similar, life. It would also require less building materials than an AU-scale Dyson Sphere, still provide $O(10^{5})$ times the living area of a planet, and obviously not require artificial gravity technology. Such a Dyson Sphere would also radiate in the IR, but since it will have white-dwarf power$^{3}$, it would be harder to detect.

In the next section, we briefly review white dwarfs as one of the possible endstates of stellar evolution, and discuss why it is reasonable to suppose that some of them might also have Dyson Spheres. In section 3,  we calculate the relation between the temperature and gravity on the Dyson Sphere in terms of the power$^{3}$ and mass of the central object. We then draw temperature-gravity graphs for a representative sample of white dwarfs, and observe that for about one third of the white dwarfs, acceptable radii for Dyson Spheres can be found. In section 4, we discuss some details such as the material requirements, the stresses, and the potential of using waste materials to create more energy.
In the final section, we conclude.


\section{White Dwarfs, Life and Dyson Spheres }
\label{WD}

\begin{figure}[h!]
\begin{center}
\includegraphics[scale=0.55]{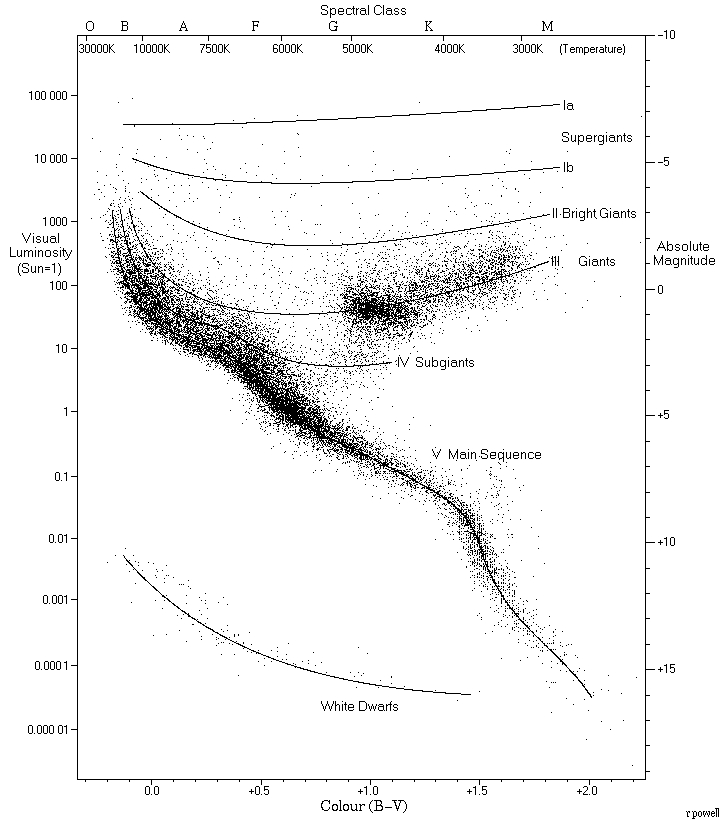}
\end{center}
\caption{The Hertzsprung-Russel (HR) diagram for a representative sample of stars; 22000 stars from the Hipparcos Catalogue together with 1000 low-luminosity stars (red and white dwarfs) from the Gliese Catalogue of Nearby Stars. \footnotesize {\it Adapted from Richard Powell via Wikipedia.}}
\label{fig:HR}
\end{figure}

   It is well-known (see any introductory astronomy textbook) that stars are classified by using the so-called Hertzprung-Russel (HR) diagram, where each star is represented as a point in the absolute magnitude-spectral class (i.e. Luminosity-Surface Temperature) plane. The stars are not distributed uniformly on that plane, but come in three main groups (Fig. \ref{fig:HR}). The diagonal band is called the main sequence, the group at the lower left are the white dwarfs, and the group roughly upper right are red giants. The names are appropriate: temperature decreases to the right, and high temperature stars are blue-white, low temperature stars are red. Also, the luminosity of a star in terms of its radius, and surface temperature is given by  
\begin{equation}
L=4 \pi R^2 \sigma T^4, \label{LRT}
\end{equation}
so we have $ R\propto\frac{\sqrt{L}}{T^2} $, therefore radius increases from the lower-left to upper-right.

   It turns out that these three groups represent evolutionary stages in the life of a star. The main sequence consists of stars fusing hydrogen, whereas the red giants are stars who have ended their main-sequence life and now are fusing higher elements at furious rates, white dwarfs are lower mass stars that have exhausted the possible fusion reactions and are slowly cooling by radiation. The white dwarfs represent one of the three possible end-states of stellar evolution, the other two being neutron stars and black holes. 

   Stars with masses of up to approximately 4 solar masses will eventually become white dwarfs. The limit is uncertain, since stars eject some mass into space during the red giant stage, and the criterion is the mass that remains. Even then, rotation makes a difference, but a non-rotating star will turn into a white dwarf if the remaining mass is less than the Chandrasekhar mass limit which is 1.4 solar masses.

   Presumably, conditions should stay consistent for a relatively long time for life to flourish and diversify; extreme deviations result in mass extinctions. Also presumably, intelligence arises only after a long process of evolution. Since the main-sequence lifetime of a star is roughly proportional to the inverse cube of its mass, one would expect the stars of most planetary systems harboring intelligent life to eventually end up as white dwarfs.  If interstellar travel is fundamentally problematic, e.g. due to energy requirements --kinetic energy of a relativistic spaceship must be several, maybe tens of times its rest energy-- or the vastness of interstellar distances implying one way or generations-long trips, building a Dyson Sphere around the newly formed white dwarf might be the natural way of sustaining the existence of the civilization.
   
Actually, the red giant stage is also quite long, about a billion years for a solar-mass star (it decreases with mass), so it might seem that life/intelligence/civilization could also develop during that stage. However, conditions like radiative flux and stellar wind, even the mass of the star, are quite variable during the red giant phase, as opposed to the main-sequence period, when they are stable. 
Hence, while the astronomically-long-term outlook of an intelligent civilization is subject of conjecture \cite{dick}, a civilization set to build a Dyson Sphere as discussed in this work would probably arise during the main sequence period of its star, and find ways to survive the red giant stage; maybe temporarily migrating to an orbit farther from its star, either to a planet, or failing/rejecting\footnote{
The reader may want to look up the concept of "Planetary chauvinism", coined by Isaac Asimov, apparently originally due Carl Sagan.}
 that, to orbital habitats a la O'Neill \cite{oneill}. In fact, such an effort would provide the experience needed before undertaking the construction of the Dyson Sphere around the eventual white dwarf.


\section{Analysis of suitability of Dyson Spheres around White Dwarfs for human life} \label{analysis}

  The temperature and the gravitational field on a Dyson Sphere are both functions of its radius, once the mass and luminosity of the central object are given. Therefore, on the temperature-gravity plane, all  possible Dyson Spheres around a given white dwarf will fall on a specific curve, the radius of the sphere being a parameter along the curve. On the same plane, the acceptable ranges for Earth-like life will define a rough rectangle, and we are interested in how many curves (each corresponding to a certain white dwarf) pass through that rectangle.  
    
   The gravity on the Dyson Sphere of radius $r$ built around a white dwarf  of mass $M$ and luminosity $L$ will be 
\begin{equation}
g=\frac{GM}{r^{2}}
\end{equation}
and the temperature
\begin{equation}
T={\left(\frac{L}{4\pi\sigma r^2}\right)}^{1/4},
\end{equation}
since in thermal equilibrium, the white dwarf's luminosity  will be emitted thermally from the outside [see equation (\ref{LRT})]. Here, obviously any energy storage ``on board'' the Dyson Sphere is ignored, because in the long term, any energy used for lighting, transportation, etc. will eventually turn into heat. We also assume that the ecology has reached a steady state; even during the population of the sphere, the energy storage rate in the chemical bonds of newly created living matter will be negligible with respect to the flux of energy from the white dwarf. Hence,
\begin{equation}
T=\left(\frac{Lg}{4\pi\sigma GM}\right)^{1/4} \label{Lg}
\end{equation}
  i.e. we need to know the luminosities and masses for the white dwarfs.

\begin{figure}[h!]
\begin{center}
\includegraphics[scale=0.35]{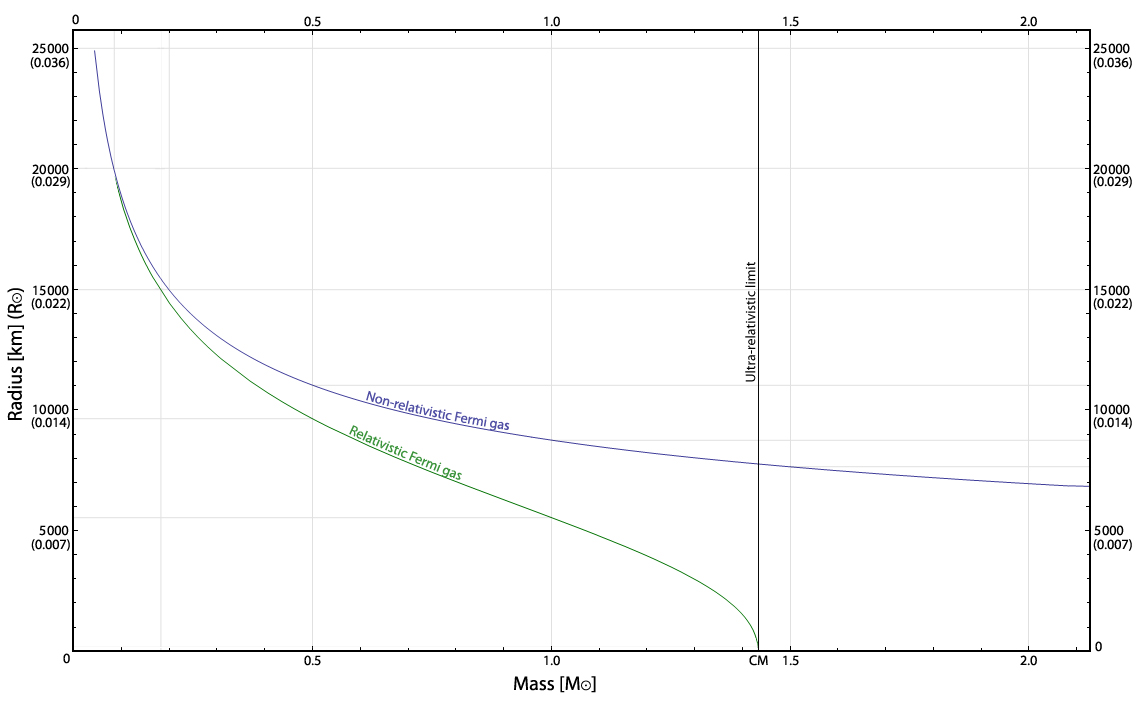}
\end{center}
\caption{Mass-radius relation for white dwarfs. \footnotesize {\it Credit: Wikimedia Commons}}
\label{fig:M-R}
\end{figure}

    While catalogues of white dwarfs exist \cite{catalogue} the information listed is not a form transparent to the potential readers of this article. Instead, we turn to an HR  diagram (see Figure \ref{fig:HR}), which shows the required information, when it exists, in pictorial form. It shows 142 white dwarfs.
    Zooming into the white dwarfs section of enlarged version of Figure \ref{fig:HR}, we carefully read off the luminosity and color index values for each white dwarf. To find the mass, we need the radius of the white dwarfs, since the mass-radius relationship is known. The radius can be found as outlined in Section 2, so we need the (effective) surface temperatures which we find from the color index formula \cite {color-index},
\begin{equation}
B-V = \left\{
\begin{tabular}{lcl}
$-3.684 \log(T) + 14.551$ & for & $\log(T) \leq 3.961$\\
$0.344 [\log(T)]^{2} -3.402 \log(T) + 8.037$ & for & $\log(T) > 3.961$
\end{tabular} \right.
\end{equation}
Then equation (\ref{LRT}) gives the radius, and from the graph of the mass-radius relationship (Figure \ref{fig:M-R}), the mass can be carefully read off.

\begin{figure}
\centering
\includegraphics[scale=0.32]{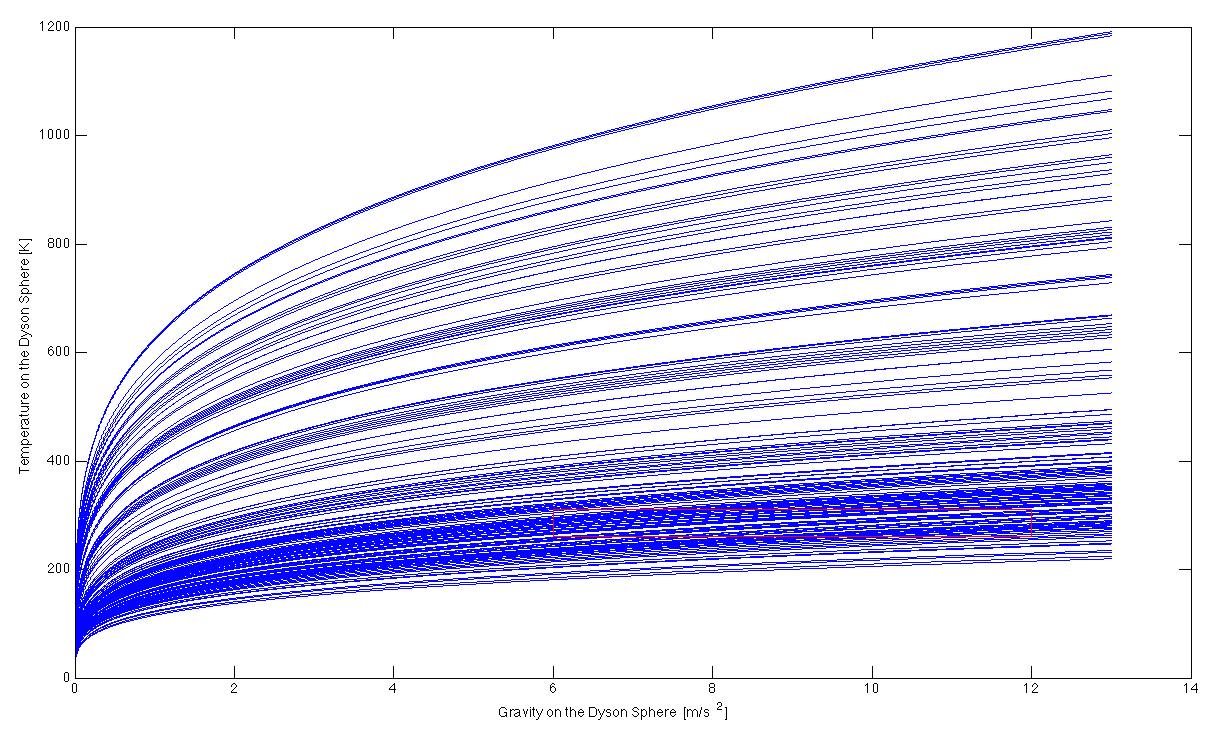}
\caption{Temperature and gravity on Dyson Spheres around white dwarfs. Each curve represents one particular white dwarf; the habitable rectangle 6-12 m/${\rm s^{2}}$ - 260-310 K is indicated.}
\label{fig:ours}
\end{figure}

   Finally, we can plot the temperature-gravity relationships for all considered white dwarfs (Figure \ref{fig:ours}).

\begin{figure}[h!]
\begin{center}
\includegraphics[scale=0.34]{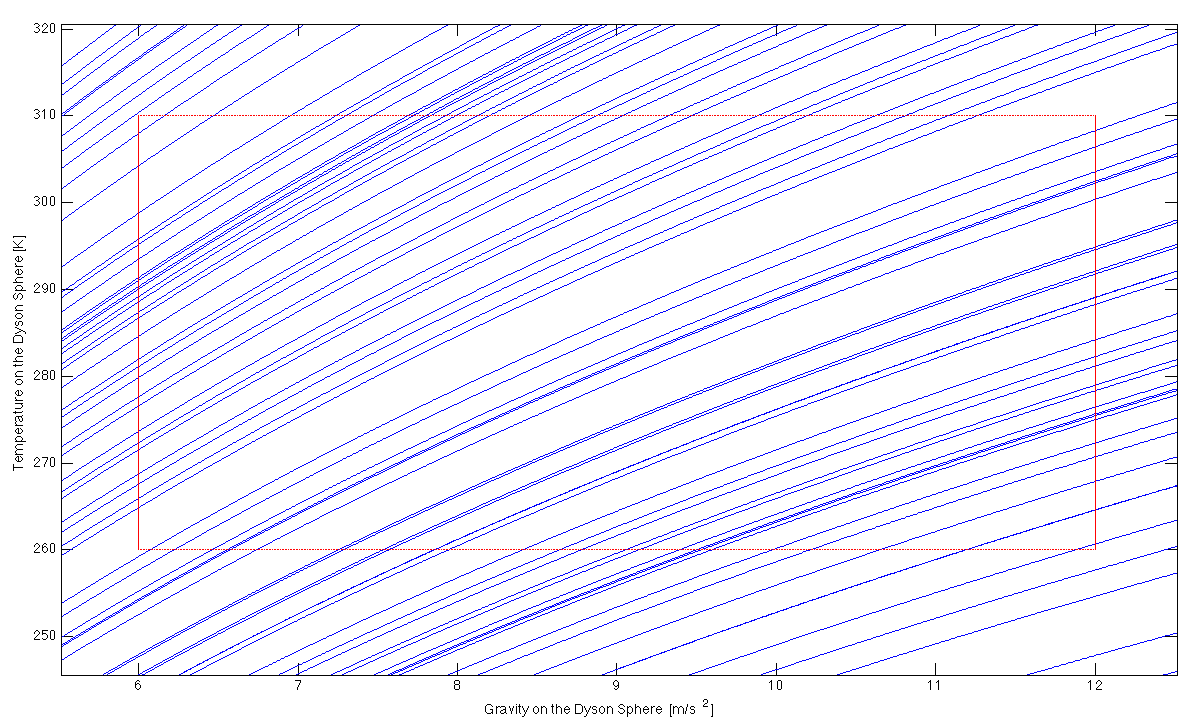}
\end{center}
\caption{Close-up of the habitable rectangle region of Fig. \ref{fig:ours}. Again, each curve represents one white dwarf.}
\label{fig:habitableRectangle}
\end{figure}

\begin{figure}[h!]
\begin{center}
\includegraphics[scale=0.35]{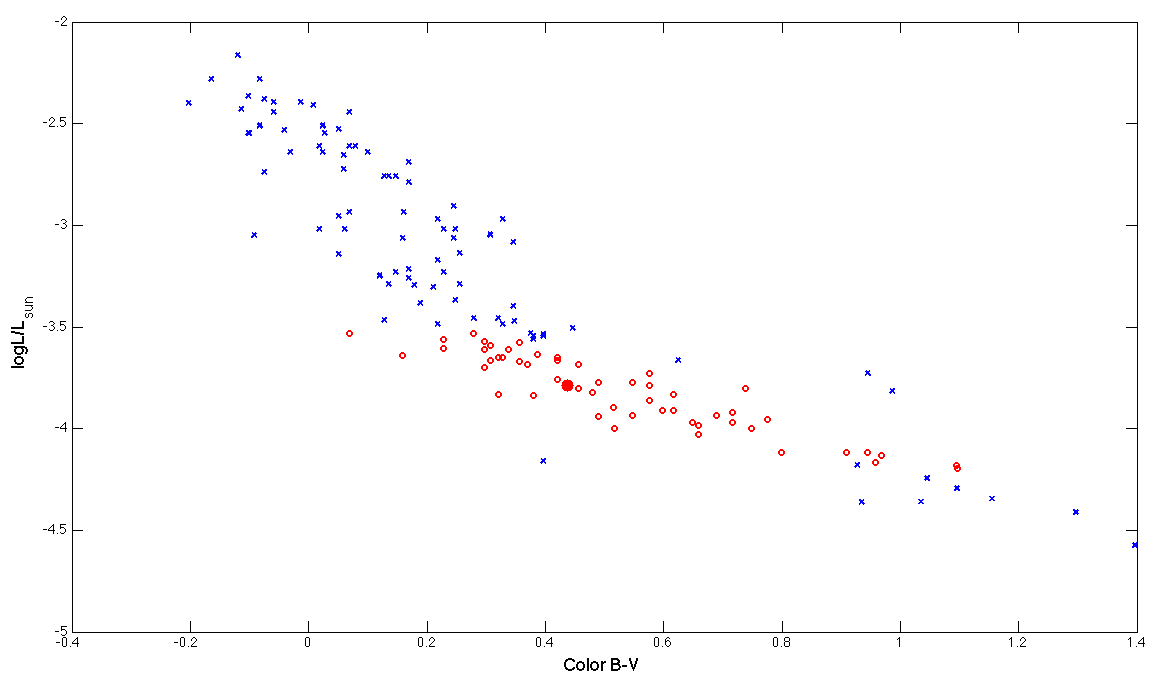}
\end{center}
\caption{The suitable and unsuitable White Dwarfs shown on the HR Diagram. The suitable ones are those whose temperature and gravity values are in the acceptable rectangle (see Fig. \ref{fig:habitableRectangle}), and are shown by red `o's; the unsuitable ones by blue crosses. The `most suitable' one, the one that crosses the rectangle of Fig. \ref{fig:habitableRectangle} diagonally, is highlighted.}
\label{fig:WDsClassified}
\end{figure}

The habitable rectangle is chosen as \mbox{$6 \; {\rm m}/{\rm s}^{2} \leq g\leq  12 \; {\rm m}/{\rm s}^{2}$} and \mbox{260 K $\leq T \leq$ 310 K}. Then, we see that the curves for 55 of the 142 white dwarfs in our set pass through that region (Figure \ref{fig:habitableRectangle}), a good ratio. Figure \ref{fig:WDsClassified} shows the suitable and unsuitable white dwarfs on the HR diagram, and the properties of suitable white dwarfs and their possible Dyson Spheres are summarized in Table \ref{table:summary}.
\begin{table}[h!]
 \begin{tabular}{ | p{30 mm} || p{10 mm}  | p{19 mm} | p{13 mm} || p{13 mm} | p{22 mm} | p{13 mm} |} \hline
\multirow{2}{*}{{\bf Description}} & \multicolumn{3}{c||}{{\bf White Dwarf}} & \multicolumn{3}{c|}{{\bf Dyson Sphere}} \\  \cline{2-7}

& Mass ($M_{\odot}$) & Luminosity ($L_{\odot}$) & Radius (km)  & Radius (km) & Temperature (K) & Gravity (${\rm m}/{\rm s}^2$)\\  \hline \hline

Most suitable WD 
 & \multicolumn{1}{c|}{0.88} & 1.6 $\times 10^{-4}$ & 6.5 $\times 10^{3}$ & 3.1-4.4 $\times 10^{6}$ & \multicolumn{1}{c|}{308-260} & 12-6.1 \\  \hline

Most massive suitable WD 
 & \multicolumn{1}{c|}{1.03} & 1.5 $\times 10^{-4}$ & $ 5.3 \times 10^{3}$ & 3.4-4.2 $\times 10^{6}$ & \multicolumn{1}{c|}{289-260} & 12-7.9   \\  \hline

Most luminous suitable WD A 
& \multicolumn{1}{c|}{0.80} & 2.9 $\times 10^{-4}$ & 7.1 $\times 10^{3}$  & 4.1-4.2 $\times 10^{6}$ & \multicolumn{1}{c|}{310-308} & 6.2-6.0 \\  \hline

Most luminous suitable WD B 
& \multicolumn{1}{c|}{1.01} & 2.9 $\times 10^{-4}$ & 5.5 $\times 10^{3}$  & 4.1-4.7 $\times 10^{6}$ & \multicolumn{1}{c|}{310-290} & 7.8-6.0 \\  \hline

System with largest possible habitable DS 
 & \multicolumn{1}{c|}{1.02} & 2.3 $\times 10^{-4}$ & 5.4 $\times 10^{3}$ & 3.6-4.8 $\times 10^{6}$ & \multicolumn{1}{c|}{310-272} & 10.2-6.0\\  \hline 

System with smallest possible habitable DS, containing the least massive and least luminous suitable WD 
 & \multicolumn{1}{c|}{0.34} & 3.8 $\times 10^{-5}$ & $11.7 \times 10^{3}$ & 1.9-2.1 $\times 10^{6}$  &  \multicolumn{1}{c|}{271-260} & 12-10.2 \\  \hline

\end{tabular}
\caption{Properties of some particular suitable white dwarfs and the associated potential Dyson Sphere ranges. The radius, temperature and gravity data are given from the smallest to the largest suitable Dyson Sphere for a given white dwarf. The most suitable white dwarf is highlighted in Figure \ref{fig:WDsClassified}.} 
\label{table:summary}
\end{table}


\section{Further Considerations} \label{more}

Denizens of the Dyson Sphere would live on the outside of the sphere, using energy collected on the inside surface, e.g. photovoltaically. This means that they will either have to use artificial lighting, or light pipes. Both possibilities will facilitate creation of day/night cycles, if the metabolisms of the denizens requires it, as expected for creatures originating on a rotating planet. They could also use nuclear power (fission or fusion) for extra energy, which will increase the infrared emitted out by the Dyson Sphere slightly.

\subsection{Amount of Construction Material}

The mass required for the construction of the Dyson Sphere is
\begin{equation}
M = \rho4 \pi r^2 t \label{massDS}
\end{equation}
where $\rho$ is the density of the material of the Sphere and $t$ its thickness. For Earth-like density, radius of $3 \times 10^{6}$ km and thickness of 1 meter, we find a mass of $\sim 6 \times 10^{23}$ kg, slightly less than the mass of the Moon! Obviously this is a small fraction of the usable mass in the solar system.\cite{Sandberg}, and the mass of one  terrestrial planet will easily give a 10 m-thick shell, so that the inhabitants will not worry very much about accidentally puncturing it.  

\subsection{Energy from Trash}

The white dwarf would also serve as a trash-to energy converter for the denizens of the Dyson Sphere: Any object dumped --or released-- from the  Sphere would fall straight onto the white dwarf, which of course is also a very good reason for {\em not} living on the inside. This material  would be forever lost to the Sphere, unlike the trash on a planet. Since in a complex ecology, one creature's trash is often another's food, organic material should probably not be disposed of this way, except for very toxic or dangereous ones. But for materials that {\em are} toxic or dangereous, organic or otherwise, the white dwarf will be the perfect place of disposal. Moreover, the trash will hit the surface of the WD with kinetic energy approximately equal to $GMm/R$, which for a WD of Solar mass and Earth radius, will get converted into heat energy equal to $2.3 \times 10^{-4} \; mc^{2}$, comparable to energy released in fission reactions. This energy will contribute to the radiation of the WD, and because the radiation is intercepted by the Dyson Sphere, the trash has effectively given as much energy as the fuel material of a nuclear fission reactor! This high release of heat by infalling material is not surprising when one remembers that in white dwarf/red giant binaries, the hydrogen from the red giant, upon falling onto the surface of the WD, can reach fusion-igniting temperatures, precipitating the catacylismic event called a {\em nova explosion}. However, since trash will probably not be dumped at a rate similar to accreting white dwarf/red giant binary systems, there is no reason to fear such a catastrophic event.

Another potential worry is that by adding mass to the white dwarf, one might push it over the limit of instability, setting off a {type Ia supernova}. However, the mass of the Dyson Sphere will typically be negligible with respect to that of the white dwarf, hence the WD's mass cannot be modified appreciably by the Sphere's inhabitants.

\subsection{Stricter Strength Requirements}

The required compressive strength of the material for the rigid Dyson Sphere is (see Appendix)
\begin{equation}
S = \frac{F}{A} = \frac{GM}{2r}\rho,
\label{eq:stress}
\end{equation}
interestingly, independent of thickness. Therefore, these DS's will require materials much stronger than AU-size DS's, for example, for material with Earth-like density, WD with Sun-like mass and DS radius of $3 \times 10^{6}$ km, we find a required strength of $\sim 10^{13}$ N/${\rm m}^{2}$. Such material strength is simply unattainable: We may estimate that a chemical bond could resist a force of the order of $1 \; {\rm eV}/\textrm{\AA} \approx 10^{-9}$ N, since typical bond energies are of the order of eV's and bond lengths of the order of angstroms. The unit area will have of the order of $10^{20}$ chemical bonds in the perpendicular direction, hence can support at most $\sim 10^{11}$ N, even if all the bonds are perfectly aligned without any imperfections in the molecular structure, and are able to share the load equally.

Therefore, a rigid Dyson Sphere around a white dwarf could not be built without some extra means of support, about which we choose not to speculate here.


\section{Conclusions}

We argued that most stars (if any) harboring intelligent life must end up as white dwarfs, hence it is natural to also consider these objects in addition to main sequence stars as central objects for Dyson Spheres. More importantly, it is possible to find parameters for these DS's such that both the temperature and gravity are close to the values for the type of life we are familiar with. This should be contrasted with the usual idea of the Dyson Sphere, AU-sized and around a main-sequence star, where gravity is negligible on the DS, hence artificial gravity may be needed, a technology which might be impossible.

Other benefits are much less need for building materials compared to the `standard' DS, and the possibility of converting trash into energy at nuclear-fission-reactor efficiencies. A drawback is much stricter strength requirement for the building materials.

We conclude that $10^{6}$ km-scale Dyson Spheres built around white dwarfs are at least as realistic as the `standard' ones, and possibly more probable. Unfortunately, they would also be harder to detect.

\appendix

\section{Stresses on and Stability of the rigid Dyson Sphere}

\subsection{Stress on the rigid Dyson Sphere}

\begin{figure}[h!]
\begin{center}
\includegraphics[scale=0.6]{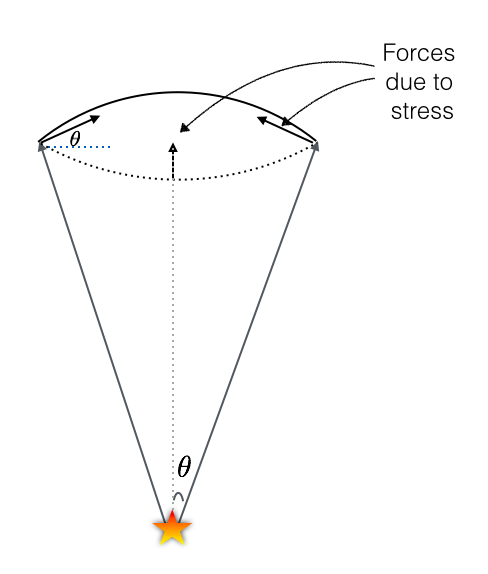}
\end{center}
\caption{Calculating the stress on a Dyson Sphere. }
\label{fig:stress}
\end{figure}

Consider a small circular section of the spherical shell, a cap, with angular radius $\theta << 1$ (Figure \ref{fig:stress}). Its mass will be $m = \pi (r\theta)^2 t \rho$, where $t$ is the thickness and $\rho$ the density of the shell. Due to the stress in the shell, forces will act on the circumference of the cap, whose outward components will add, and components in the direction perpendicular to it will cancel. The net outward force  will be $2 \pi (r\theta) T_{\rm s} \sin \theta$, where $T_{\rm s}$ is the force per unit length along the circumference (like a negative {\it surface tension}). Now, in terms of the stress $S$, we have $T_{\rm s} = S t$. Putting the gravitational force on the cap equal to the net outward force, we get
\begin{equation}
\frac{ G M \pi r^{2} \theta^{2} t \rho}{r^{2}} = 2 \pi r \theta S t \sin \theta,
\end{equation}
which, using also the approximation $\theta << 1$, easily gives eq.(\ref{eq:stress}).

\subsection{Stability -- Sphere \& Ring}

The rigid Dyson Sphere around a star is neutrally stable. The easiest way to understand this is to recall the well-known electrostatic result that the electric field inside a uniform spherical shell vanishes. Therefore, a point charge inside such a shell will not experience a force, and by Newton's third law, will not exert a net force on the shell. Since the gravitational force obeys exactly the same $1/r^{2}$ law as the electrostatic force, the same result must apply.

\begin{figure}[h!]
\begin{center}
\includegraphics[scale=0.6]{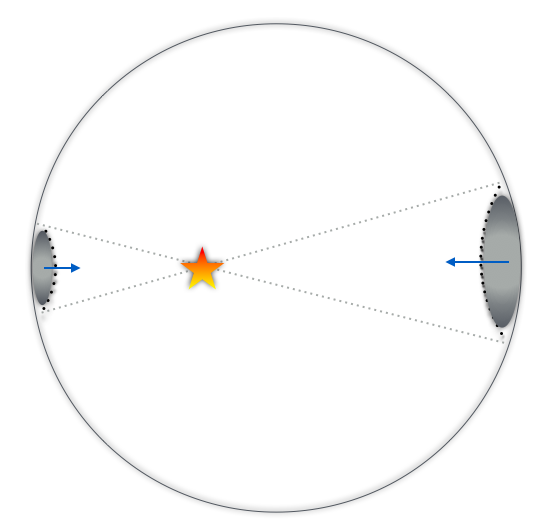}
\end{center}
\caption{Analyzing the stability of a Dyson Sphere or Niven ring.} 
\label{fig:stability}
\end{figure}

A more pictorial way to understand the result --also useful for comparison to the ring case-- is  to consider a double-cone with apex at the star, and small opening angle (Fig. \ref{fig:stability}). For the sphere, the mass cut out from the shell by one cone will be proportional to $r^{2}$, where $r$ is the distance from the star to the shell. Therefore, the gravitational force exerted on such a mass, $GMm/r^{2}$, will be independent of $r$, hence the same for the two masses cut out by the two sides of the double-cone, even if the star or white dwarf is not centered.

For a Niven-type~\cite{ringworld} ring, however, the mass cut out  by a cone is proportional to $r$, hence the gravitational force exerted on the  mass is proportional to $1/r$~! This means that the nearer side will be attracted stronger than the far side, hence any small displacement of the ring with respect to the star, originally centered, will get amplified; and eventually one side of the ring will hit the star. Of course, any civilization advanced enough to build such a structure will also see this danger and must take measures to prevent it, which is one of the themes of  the next book~\cite{RW_Eng} in the {\it Ringworld} series.

\end{document}